\documentclass[runningheads]{llncs}
\usepackage[T1]{fontenc}
\usepackage{graphicx}
\usepackage{multirow} 
\usepackage{makecell}
\usepackage{booktabs}
\usepackage{amsfonts}
\usepackage{amsmath}
\usepackage{ulem}
\usepackage{hyperref}
\usepackage{float}

\hypersetup{hidelinks}

\def\etal{\normalem\emph{et al.~}}
\def\ie{\normalem\emph{i.e.}}
\def\aka{\normalem\emph{a.k.a.}}

\newfloat{figtab}{htb}{fgtb}
\makeatletter
  \newcommand\figcaption{\def\@captype{figure}\caption}
  \newcommand\tabcaption{\def\@captype{table}\caption}
\makeatother

\begin{document}
\title{ARHNet: Adaptive Region Harmonization for Lesion-aware Augmentation to Improve Segmentation Performance}

\titlerunning{ARHNet}

\author{Jiayu Huo\inst{1} \and
Yang Liu\inst{1} \and
Xi Ouyang\inst{2} \and
Alejandro Granados\inst{1} \and \\
S\'{e}bastien Ourselin\inst{1} \and
Rachel Sparks\inst{1}
}

\authorrunning{J. Huo et al.}

\institute{School of Biomedical Engineering and Imaging Sciences (BMEIS), \\
King's College London, London, UK \\
\email{jiayu.huo@kcl.ac.uk}
\and
Shanghai United Imaging Intelligence Co., Ltd., Shanghai, China \\
}


\maketitle              
\begin{abstract}
Accurately segmenting brain lesions in MRI scans is critical for providing patients with prognoses and neurological monitoring. However, the performance of CNN-based segmentation methods is constrained by the limited training set size. Advanced data augmentation is an effective strategy to improve the model's robustness. However, they often introduce intensity disparities between foreground and background areas and boundary artifacts, which weakens the effectiveness of such strategies. In this paper, we propose a foreground harmonization framework (ARHNet) to tackle intensity disparities and make synthetic images look more realistic. In particular, we propose an Adaptive Region Harmonization (ARH) module to dynamically align foreground feature maps to the background with an attention mechanism. We demonstrate the efficacy of our method in improving the segmentation performance using real and synthetic images. Experimental results on the ATLAS 2.0 dataset show that ARHNet outperforms other methods for image harmonization tasks, and boosts the down-stream segmentation performance. Our code is publicly available at \url{https://github.com/King-HAW/ARHNet}. 
\end{abstract}

\section{Introduction}
Accurate brain lesion segmentation is essential for understanding the prognoses of neurological disorders and quantifying affected brain areas by providing information on the location and shape of lesions~\cite{liew2022large}. With advanced deep learning techniques, various brain lesion segmentation methods based on Convolutional Neural Networks (CNNs) have been proposed~\cite{liu2019msdf,zhang2020mi}. However, a noteworthy hurdle is the prerequisite of an adequate number of training samples to ensure the model's generalization ability. Utilizing small-scale datasets for the segmentation model training can result in over-fitting, thereby limiting its robustness to unseen samples. Due to the variance of lesion appearance and size, as well as the extreme data imbalance between foreground and background voxels, many deep learning models also struggle to perform the small lesion segmentation task.

To this end, some data augmentation techniques have been proposed that aim to increase the diversity of the training set, which helps to boost the performance of the segmentation model for unseen images~\cite{chen2020group}. Often data augmentation is realized by basic image transformations such as rotation and flipping. As the diversity of the data generated through basic image transformations is deficient, advanced data augmentation approaches have been developed. For instance, Huo~\etal~\cite{huo2022brain} designed a progressive generative framework to synthesize brain lesions that can be inserted into normal brain scans to create new training instances. Zhang~\etal~\cite{zhang2021carvemix} proposed a lesion-aware data augmentation strategy to increase the sample diversity. However, these methods often inevitably introduce boundary artifacts that may cause the intensity distribution to shift, resulting in segmentation performance degradation~\cite{wei2018adversarial}. Recently, some image harmonization frameworks~\cite{cong2020dovenet,ling2021region} have been developed to solve the boundary and style discontinuities between the foreground and background for natural images. However, these frameworks have limitations when applied to brain MRI scans, where the smooth transition between the lesion and surrounding tissues is more critical than natural images.

In this paper, we tackle the problem of foreground intensity and style mismatch created by data augmentation, so that plausible images can be generated. As we do not have paired real and simulated images, we create simulated images by taking real images and introducing foreground disparities to use for training the image harmonization network (ARHNet). We further present an Adaptive Region Harmonization (ARH) module to align foreground feature maps guided by the background style information. Finally, we train a segmentation model based on the mixture of real and synthetic images produced by ARHNet to demonstrate its effectiveness for improving down-stream segmentation performance.

\section{Methodology}
The purpose of ARHNet is to harmonize the foreground in augmented images created by a data augmentation technique such as Copy-Paste~\cite{ghiasi2021simple}, to further serve downstream tasks like segmentation. We try to find a function $f$ such that $f_{\boldsymbol{\theta}} (\tilde{I}_a, M_a) \approx I_a$. Here, $\tilde{I}_a$ is the augmented image, $I_a$ is the corresponding real image, and $M_a$ is the foreground mask of $\tilde{I_a}$. $\boldsymbol{\theta}$ refers to the parameter vector of $f$, \aka, ARHNet. However, since the augmented image $\tilde{I}_a$ does not have a corresponding real image $I_a$, we perform foreground intensity perturbation using a real brain MRI scan $I$ with stroke lesions and its foreground mask $M$ to create an image $\tilde{I}$ that simulates $\tilde{I}_a$ with a disharmonious foreground. 
We train ARHNet using the pairs $(\tilde{I}, M) \rightarrow I$ to learn the parameter vector $\boldsymbol{\theta}$.

\subsection{Overview of ARHNet}
\label{section_framework_overview}
\begin{figure}[!t]
\centering
\includegraphics[width=0.98\textwidth]{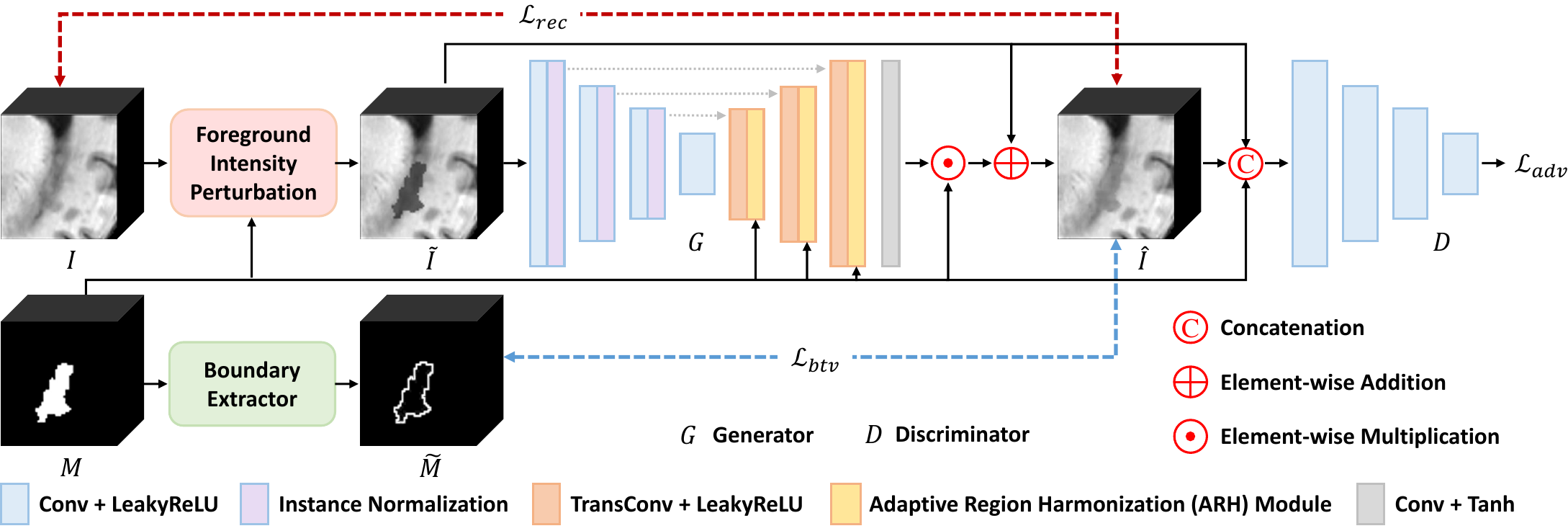}
\caption{The pipeline of ARHNet for adaptive image harmonization for simulated brain MRI with stroke lesions. 
} 
\label{fig:main_framework}
\end{figure}

Fig.~\ref{fig:main_framework} represents our framework (ARHNet) for foreground harmonization, which comprises four components: a foreground intensity perturbation unit, a boundary extractor, a generator $G$, and a discriminator $D$. 
Given $I$ and $M$, 
$I$ is first scaled from 0 to 1. Next, the foreground intensity perturbation unit generates a foreground intensity-perturbed image $\tilde{I}$. Intensity perturbation is performed as follows:
\begin{equation}
\tilde{I}=\left[ {\left( {1 + \alpha } \right) \cdot I + \lambda } \right] \odot M + I \odot \left( {1 - M} \right),
\label{eq:intensity_perturbation}
\end{equation}
where $\alpha \sim \mathcal{U}(-0.3,0.3)$, $\lambda \sim \mathcal{U}(-0.3,0.3)$. Here $\alpha$ and $\lambda$ can simulate large intensity variance in augmented images $\tilde{I}_a$ generated by advanced data augmentation approaches like Copy-Paste~\cite{ghiasi2021simple}. ``$\odot$'' denotes element-wise multiplication. After the foreground intensity perturbation, the stroke area is either brighter or darker compared to the background tissue, which is a boundary mismatch. Next, $\tilde{I}$ is passed through $G$ to obtain the intensity difference map. The foreground region of the intensity difference map is then extracted using $M$ and further added by $\tilde{I}$ to get a harmonized image $\hat{I}$. Inspired by~\cite{ouyang2018pedestrian}, we concatenate $\hat{I}$ with $\tilde{I}$ and $M$ to create the input image pair for $D$. Here $\tilde{I}$ and $M$ provide location information of the foreground, which benefits the adversarial training process and ensures $\hat{I}$ have high fidelity to the ground truth image.

To optimize $G$ and $D$ so that harmonized images $\hat{I}$ have realistic texture and harmonized boundary intensities, three loss functions are deployed during model training: reconstruction loss $\mathcal{L}_{rec}$, boundary-aware total variation loss $\mathcal{L}_{btv}$, and adversarial loss $\mathcal{L}_{adv}$. The reconstruction loss implemented in our framework is defined as:
\begin{equation}
\mathcal{L}_{rec}=\|I-\hat{I}\|_1.
\label{eq:loss_rec}
\end{equation}
Reconstruction L1 loss makes the output and ground truth have similar appearances but may cause over-smoothing of images. Therefore, the model tends to output images with low mean square error but with relatively blurred texture. To prevent texture blurring we add a discriminator so that the generator will produce distinct and realistic images. The adversarial loss $\mathcal{L}_{adv}$ is added as additional supervision to the training process. In particular, we use hinge loss~\cite{lim2017geometric} instead of the cross-entropy loss to stabilize the training process and prevent the gradient from vanishing. The $\mathcal{L}_{adv}$ is formulated as follows:
\begin{equation}
\mathcal{L}_{adv}(D) = \mathbb{E}_{\hat{I},\tilde{I},M}[max(0,1-D(\hat{I},\tilde{I},M))]+\mathbb{E}_{I,\tilde{I},M}[max(0,1+D(I,\tilde{I},M))],
\label{eq:loss_d}
\end{equation}
\begin{equation}
\mathcal{L}_{adv}(G) = - \mathbb{E}_{\hat{I},\tilde{I},M}[D(\hat{I},\tilde{I},M)].
\label{eq:loss_g}
\end{equation}
A loss with only $\mathcal{L}_{rec}$ and $\mathcal{L}_{adv}$ leads to an abrupt boundary between the foreground and background. To encourage the network to give low gradients on the border area of $\hat{I}$ and make the transition from background to foreground smoother, we present a boundary-aware total variation loss $\mathcal{L}_{btv}$. If $\tilde{M}$ is the set of boundary voxels extracted by the boundary extractor, $\mathcal{L}_{btv}$ can be defined as:
\begin{equation}
\mathcal{L}_{btv}=\sum_{(i, j, k) \in \tilde{M}}\|\hat{I}_{i+1, j, k}-\hat{I}_{i, j, k}\|_1+\|\hat{I}_{i, j+1, k}-\hat{I}_{i, j, k}\|_1+\|\hat{I}_{i, j, k+1}-\hat{I}_{i, j, k}\|_1,
\label{eq:loss_btv}
\end{equation}
where $i$, $j$ and $k$ represent the $(i,j,k)^{th}$ voxel in $\tilde{M}$. By adding the boundary-aware loss, our framework makes the boundary transition smoother compared to other methods (see Fig.~\ref{fig:qualitative_results_on_testset} and~\ref{fig:qualitative_results_on_real_images}), which makes the harmonized images more like those observed on real MRI. 
Overall, our total loss function is defined as:
\begin{equation}
\mathcal{L}_{total} = \lambda_{rec}\mathcal{L}_{rec}+\lambda_{btv}\mathcal{L}_{btv}+\lambda_{adv}\mathcal{L}_{adv},
\label{eq:loss_total}
\end{equation}
where $\lambda_{rec}$, $\lambda_{btv}$ and $\lambda_{adv}$ are weighting factors for each term.

\subsection{Adaptive Region Harmonization (ARH) Module}
\label{section_adaptive_region_harmonization_module}
\begin{figure}[!t]
\centering
\includegraphics[width=0.98\textwidth]{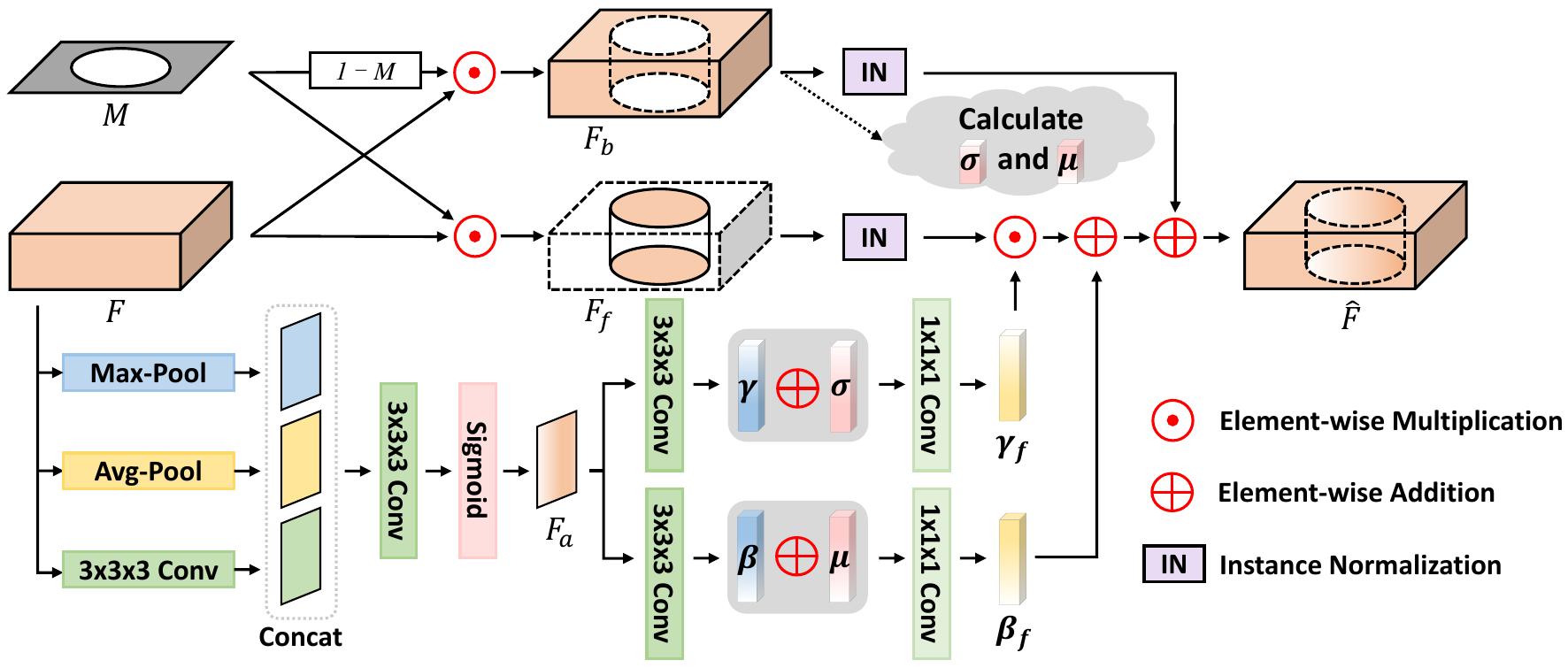}
\caption{The structure of our Adaptive Region Harmonization (ARH) module. $\mu$ and $\sigma$ represent the channel-wise mean value and standard deviation calculated from $F_{b}$.} 
\label{fig:arh_module}
\end{figure}

To better align the foreground and background feature maps obtained from $\tilde{I}$, we design a new feature normalization paradigm called Adaptive Region Harmonization (ARH) module. As depicted in Fig.~\ref{fig:arh_module}, the ARH module takes the resized foreground mask $M$ and the feature maps $F$ as input. Here $F \in \mathbb{R}^{C \times H \times W \times D}$ and $M \in \mathbb{R}^{1 \times H \times W \times D}$, where $C$, $H$, $W$, $D$ indicate the number of feature channels, height, width, and depth of $F$, respectively. We first divide the feature maps into foreground $F_{f}=F \odot M$ and background $F_{b}=F \odot (1-M)$ according to $M$. Then we normalize $F_{f}$ and $F_{b}$ using Instance Normalization (IN)~\cite{ulyanov2016instance}, and calculate the channel-wise background mean value $\mu \in \mathbb{R}^{C}$ and standard deviation $\sigma \in \mathbb{R}^{C}$ as follows:
\begin{equation}
\mu = \frac{1}{{sum(1-M)}}\sum\limits_{h,w,d} {{F_{c,h,w,d}} \odot {(1-M_{h,w,d})}},
\label{eq:mu}
\end{equation}
\begin{equation}
\sigma = \sqrt {\frac{1}{{sum(1-M)}}\sum\limits_{h,w,d} {[ {{F_{c,h,w,d}} \odot {(1-M_{h,w,d})} - \mu }]^2}},
\label{eq:sigma}
\end{equation}
where $sum(\cdot)$ indicates summing all elements in the map. Different from the RAIN module~\cite{ling2021region} that directly applies $\mu$ and $\sigma$ to $F_{f}$ to align the foreground to the background, we present a learned scaling parameter strategy, with an attention mechanism so that the network focuses more on task-relevant areas to better learn the consistent feature representation for both foreground and background. 

Specifically, we calculate an attention map $F_{a} \in \mathbb{R}^{1 \times H \times W \times D}$ based on the entire feature maps in the ARH module, to let the module adaptively extract style information from important areas. $F_{a}$ is formulated as: 
\begin{equation}
{F_a} = S(Conv([{F_{max}},{F_{avg}},{F_{Conv}}])),
\label{eq:attention_map}
\end{equation}
where $S$ denotes the sigmoid function and $Conv$ denotes the convolution operation. Additionally, we calculate two channel-wised scaling parameters $\gamma \in \mathbb{R}^{C \times H \times W \times D}$ and $\beta \in \mathbb{R}^{C \times H \times W \times D}$ as:
\begin{equation}
{\gamma} = Conv(F_a),{\beta} = Conv(F_a).
\label{eq:gamma_beta}
\end{equation}
$\gamma$ and $\beta$ allow element-wise adjustments on $\sigma$ and $\mu$ which represent the global intensity information extracted from the background feature maps. We fuse $\gamma$ and $\beta$ with $\sigma$ and $\mu$ with two convolutional layers to obtain the foreground scaling factors $\gamma_{f}$ and $\beta_{f}$, which can be calculated as:
\begin{equation}
{\gamma_{f}} = Conv(\gamma + \sigma),{\beta_{f}} = Conv(\beta + \mu).
\label{eq:gamma_f_beta_f}
\end{equation}
By applying $\gamma_{f}$ and $\beta_{f}$ to the foreground feature maps $F_{f}$, we finally attain the aligned feature maps via $\hat{F} = F_{f} \odot (1+\gamma_{f}) + \beta_{f} + F_{b}$.

\section{Experiments}
\subsection{Experiment Settings}
\subsubsection{Dataset}
We use the ATLAS v2.0 dataset~\cite{liew2022large} to evaluate the performance of ARHNet. ATLAS (short for ATLAS v2.0) is a large stroke dataset, which contains 655 T1-weighted brain MRIs with publicly available voxel-wise annotations. All images were registered to the MNI-152 template with a voxel spacing of $1mm \times 1mm \times 1mm$. According to~\cite{huo2022mapping}, about half of the images are characterized as small lesion images (foreground voxels $\leq$ 5,000). In this work we focus on only these images, corresponding to 320 MRIs. We split the dataset into five folds, stratified by lesion size to ensure both training and testing sets have the same data distribution. We randomly select one fold (20\%) as the test set and the remaining four folds are the training set. 

\subsubsection{Implementation Details}
ARHNet is implemented within PyTorch~\cite{paszke2019pytorch} and uses TorchIO~\cite{perez2021torchio} for loading data and creating intensity perturbations. To optimize the generator and discriminator, we use two AdamW optimizers~\cite{loshchilov2017decoupled}. The initial learning rates for $G$ and $D$ are set to 1$e$-4, and 5$e$-5, respectively. The batch size is set to $16$ and total training epochs are $200$ for each model. For input images, we randomly extract a $64 \times 64 \times 64$ patch from the MRI scans corresponding to the region that contains the stroke annotation(s). The loss weight factors $\lambda_{rec}$, $\lambda_{btv}$, and $\lambda_{adv}$ are set to 100, 10, 1, respectively. For the down-stream segmentation task that is used to evaluate our framework, we implement a segmentation model based on Attention UNet~\cite{oktay2018attention} in the MONAI framework~\cite{cardoso2022monai}. The initial learning rate is set to 1$e$-3, and the batch size is $4$. For a fair comparison, we train each setting for $30,000$ iterations.

\subsubsection{Evaluation Metrics}
We evaluate the performance of ARHNet on the image harmonization task and also a down-stream stroke segmentation task. For the image harmonization task, we use four metrics to measure the fidelity of the output, \ie, mean absolute error (MAE), mean absolute error of the foreground region (fMAE), peak signal-to-noise ratio (PSNR), and signal-to-noise ratio of the foreground region (fPSNR). For the down-stream stroke segmentation task, we use three metrics to evaluate the segmentation performance: the Dice coefficient, $95\%$ Hausdorff Distance (95HD), and average surface distance (ASD). 

\subsection{Experimental Results}
\subsubsection{Comparison of Image Harmonization Results}
\begin{table}[!t]
\caption{Metrics for image harmonization on the ATLAS test set. The best results are highlighted in bold. fMAE and fPSNR are computed in the foreground.}
\label{tab:quantitative_results_on_testset}
\centering
\setlength{\tabcolsep}{4.5mm}{
\begin{tabular}{rcccc}
\toprule[1.5pt]

Method & MAE$\downarrow$ & fMAE$\downarrow$ & PSNR$\uparrow$ & fPSNR$\uparrow$ \\

\hline
Composite & 0.0014 & 0.23 & 39.70 & 14.62 \\
HM        & 0.0010 & 0.14 & 41.38 & 16.30 \\
\hline
UNet~\cite{ronneberger2015u}      & 0.0009 & 0.11 & 43.94 & 18.88 \\
Hinge-GAN~\cite{lim2017geometric} & 0.0011 & 0.14 & 43.44 & 18.38 \\
UNet-GAN~\cite{ling2021region} & 0.0009 & 0.12 & 44.16 & 19.11 \\
RainNet~\cite{ling2021region} & 0.0007 & 0.09 & 45.33 & 20.30 \\
Ours      & \textbf{0.0006} & \textbf{0.07} & \textbf{46.74} & \textbf{21.74} \\

\bottomrule[1.5pt]
\end{tabular}}
\end{table}
We quantitatively compare the foreground image harmonization results of ARHNet on the ATLAS test set with other non-learning- and learning-based methods. Results are shown in Table~\ref{tab:quantitative_results_on_testset} where ``Composite'' means we do not use any image harmonization method but directly calculating the metrics based on the images with foreground disparities which are inputs for all other methods. It gives the worst results as expected.
If we adapt the foreground intensity to be consistent with the background based on Histogram Matching (``HM'' in Table~\ref{tab:quantitative_results_on_testset}), we can achieve better results, but still worse than all of the learning-based methods evaluated.

Four learning-based methods are implemented as comparisons. Here ``UNet'' refers to the UNet model trained with only the reconstruction loss $\mathcal{L}_{rec}$. ``Hinge-GAN'' means the UNet model trained with only the adversarial loss $\mathcal{L}_{adv}$. ``UNet-GAN'' denotes the UNet model is trained under the supervision of $\mathcal{L}_{rec}$ and $\mathcal{L}_{adv}$. ``RainNet'' is a generator that consists of the RAIN module~\cite{ling2021region}, also only $\mathcal{L}_{rec}$ and $\mathcal{L}_{adv}$ are used for backpropagation. From Table~\ref{tab:quantitative_results_on_testset}, we can find that our method outperforms other methods on all metrics, proving the efficacy and rationality of ARHNet. Furthermore, compared with RainNet, our method achieve a big improvement of 1.41 dB in PSNR and 1.44 dB in fPSNR.

\begin{figure}[!t]
\centering
\includegraphics[width=0.98\textwidth]{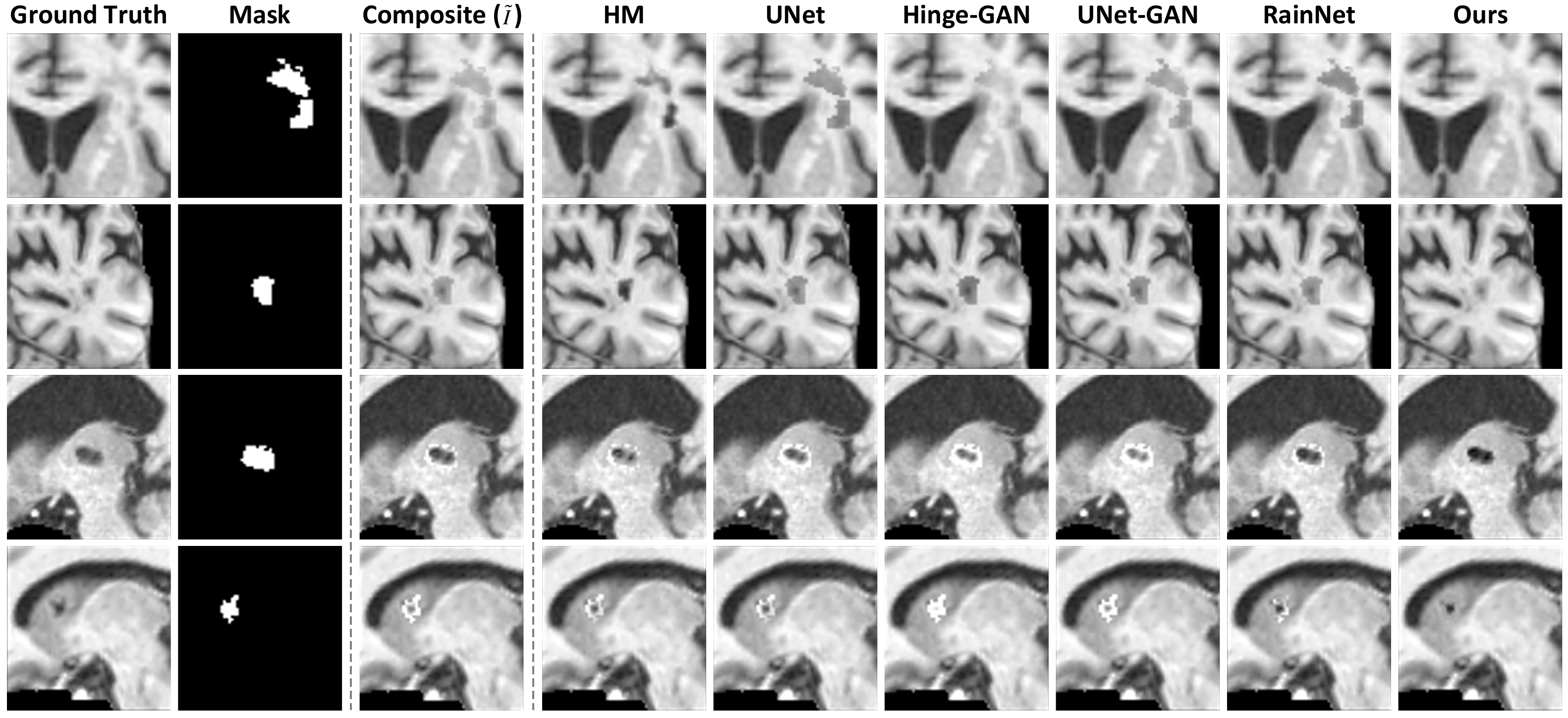}
\caption{Qualitative comparison between different harmonization methods.} 
\label{fig:qualitative_results_on_testset}
\end{figure}

\begin{figure}[!t]
\centering
\includegraphics[width=0.98\textwidth]{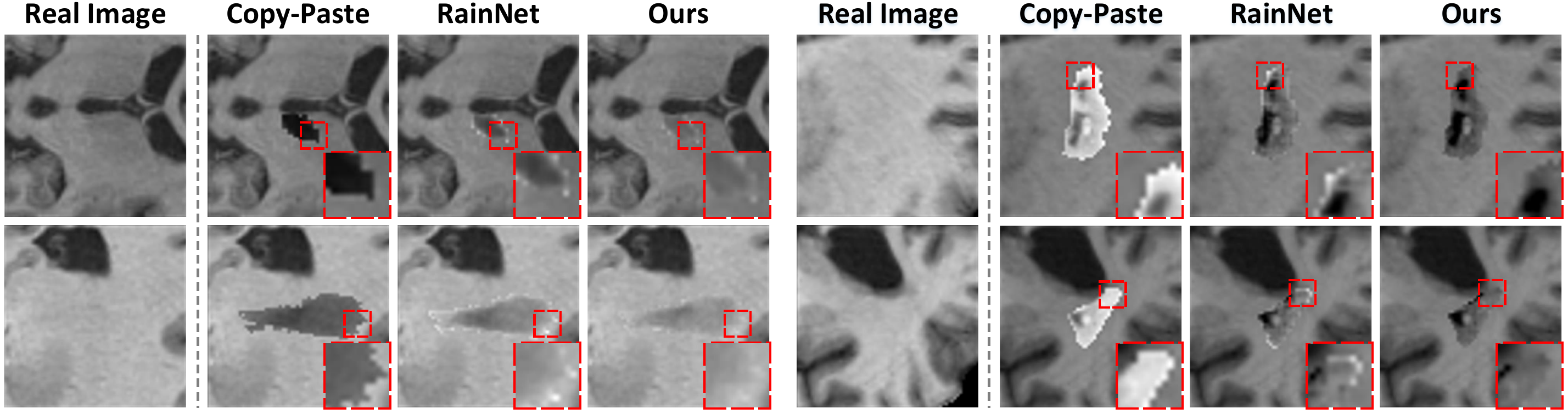}
\caption{Visualization results on composite brain MRI scans which are used for the down-steam segmentation task.} 
\label{fig:qualitative_results_on_real_images}
\end{figure}

We present qualitative results in Fig.~\ref{fig:qualitative_results_on_testset} and~\ref{fig:qualitative_results_on_real_images}. In Fig.~\ref{fig:qualitative_results_on_testset} we can observe that ARHNet can achieve realistic harmonization images no matter if the foreground is brighter or darker than the background (top two rows: darker, bottom two rows: brighter). Also, the boundaries in our results are smoother than other methods. Additionally, we show the image harmonization results on composite brain MRI scans in Fig.~\ref{fig:qualitative_results_on_real_images}. By zooming in on the boundary area, it is easy to observe that composite images harmonized by ARHNet are more realistic than RainNet, which demonstrates the superiority of our method again.

\subsubsection{Comparison of Down-Stream Segmentation Performance}
We report quantitative measures of the down-stream lesion segmentation performance for different training sets in Table~\ref{tab:segmentation_results}. For each setting, we keep the batch size the same and train for 30,000 iterations for a fair comparison. ``-'' denotes no additional data is used for model training. ``200 real'' means 200 images with big lesions (foreground voxels $>$ 5,000) from the original ATLAS v2.0 dataset are utilized as additional training samples. ``200 by~\cite{zhang2021carvemix}'' refers to using CarveMix to generate additional 200 images for model training. ``200 by Ours'' means we first use Copy-Paste~\cite{ghiasi2021simple} strategy to create 200 composite images, then we use ARHNet to adjust the foreground intensity to harmonize the images. As shown in Table~\ref{tab:segmentation_results}, our method achieves the best segmentation result and brings a large performance gain of 12.57\% in Dice compared to not using any additional data.

\begin{figure}[!t]
\begin{minipage}{0.43\textwidth}
\tabcaption{Segmentation performances under different training data settings.}
\vspace{1em}
\label{tab:segmentation_results}
\centering
\begin{tabular}{lccc}
\toprule[1.5pt]
Additional Data & Dice$\uparrow$ & ASD$\downarrow$ & 95HD$\downarrow$ \\
\hline
-                                        & 23.84 & 48.85 & 85.67 \\
200 real                                 & 25.05 & 48.93 & 88.08 \\
200 by~\cite{zhang2021carvemix} & 32.38 & 40.11 & 77.78 \\
200 by Ours                              & \textbf{36.41} & \textbf{25.14} & \textbf{49.30} \\
\bottomrule[1.5pt]
\end{tabular}
\end{minipage}
\hspace{1em}
\begin{minipage}{0.48\textwidth}
\tabcaption{Ablation studies on different feature normalization methods.}
\vspace{1em}
\label{tab:ablation_study}
\centering
\begin{tabular}{rcccc}
\toprule[1.5pt]
Method & MAE$\downarrow$ & fMAE$\downarrow$ & PSNR$\uparrow$ & fPSNR$\uparrow$ \\
\hline
BN~\cite{ioffe2015batch} & 0.0007 & 0.09 & 45.80 & 20.79 \\
IN~\cite{ulyanov2016instance} & 0.0008 & 0.09 & 45.68 & 20.66 \\
RAIN~\cite{ling2021region} & 0.0008 & 0.10 & 43.89 & 18.87 \\
Ours & \textbf{0.0006} & \textbf{0.07} & \textbf{46.74} & \textbf{21.74} \\
\bottomrule[1.5pt]
\end{tabular}
\end{minipage}
\end{figure}

\subsubsection{Ablation Study}
We also investigate the performance gain achieved by our ARH module, results are shown in Table~\ref{tab:ablation_study}. We can find that if we keep all other settings unchanged and only replace the ARH module with InstanceNorm or BatchNorm, higher PSNR is reached compared to RainNet (see in Table~\ref{tab:quantitative_results_on_testset}). This demonstrates the effectiveness of some of the additional elements we presented in this work, such as boundary-aware total variation loss and the foreground intensity perturbation unit. However, if we replace the ARH module with the RAIN module, the result is the worst among all normalization methods. This is likely because the RAIN module only considers the entire style of the background, and therefore cannot align the foreground feature maps properly.

\section{Conclusion}
In this paper, we propose an Adaptive Region Harmonization Network (ARHNet) that can effectively harmonize a target area and make the style of foreground and background consistent in this region. This framework can be utilized to harmonize synthetic samples generated by other data augmentation methods, and make these images more realistic and natural. Harmonized augmented samples can be further utilized in down-stream segmentation tasks to improve the segmentation model's generalization ability. Extensive experimental results demonstrate that our proposed method can generate style-consistent images and is effective for segmenting small stroke lesions on T1-weighted MRI.


\bibliographystyle{splncs04}
\bibliography{books}

\end{document}